\documentclass[a4paper,10pt]{article}

\usepackage[centertags]{amsmath}
\usepackage{amsfonts}
\usepackage{a4wide}

\title{Condensed Geometry}
\author{Koustubh Kabe \\
\small{Department of Physics, Lokmanya Tilak Bhavan, 
University of Mumbai, Kalina Campus, Santacruz (East),}\\ 
\small{Mumbai - 400 098 India}
}

\begin{document}

\maketitle

\begin{abstract}
A spin (dependent) system treatment of gravity is adopted akin to the Sen-Ashtekar treatment. Time is reinserted into the space ``fluid'' at the quantum Level. This time - the Lorentzian one- is shown to be a vorticity of a ``fluid particle'' of the space and the effect is integrated over all the fluid particles to incorporate time in quantum gravity. This spacetime is viewed as a fluid of future light cones called the SU(2) dipoles of causality here in the paper.The future light cone structure is soldered internally to the new variables derived in this paper to accomodate a background free physics of quantum strings. The emergence of spacetime is shown to be a first order phase transition and that of separation of gravity from the unified field to be a second order phase transition. For the former case the cosmic time is chosen as the order parameter and for the latter case the angular momentum is chosen as the order parameter. A quantum blackhole thus nucleates at transition temperature which is the Planck temperature, $\tau_{pl}$. Then the SU(2) dipoles enable interpretation of this black hole as a gravity gauge SL(2,$\mathbb{C}$) dual of the U(1) gauge ferromagnetic phase. The usual QFT interpretation of this effect is the existence of locally Lorentzian spacetimes. 
\end{abstract}

\section{Preliminaries}

We consider the configuration space $\mathcal{C}$ of pure gravity or general relativity (GR), whose cotangent bundle is the phase space $\varPi$ of the SL(2,$\mathbb{C}$) Yang-Mills gravity gauge. Fix in this configuration space a state functional $\psi$. The gauge transformation $\psi\rightarrow\psi'$ introduces a dynamical phase and a geometric phase. This geometric phase induces a SL(2,$\mathbb{C}$) Berry connection $K_{\mu}^{i}$. As in the Sen-Ashtekar formalism, we consider this connection with Lorentzian indices $\mu\nu\rho\sigma$.... and with a Lie Algebra index 'i,j,k,l...'. The SL(2,$\mathbb{C}$) Berry curvature is then 
\begin{equation}
 G_{\mu\nu}^i= ^\pm\partial_{[\mu}K_{\nu]}^i+ig[K_{\mu},K_{\nu}]^i
\end{equation}
Each $K_{\mu}^{i}$ is a positive definite square root of the metric $g_{\mu\nu}$ as 
\begin{equation}
 g_{\mu\nu}= K_{\mu}^{i}K_{\nu}^{j}\varepsilon^i\varepsilon^j
\end{equation}
where $\varepsilon$ is the SU(2) volume form.
Later we interpret this curvature as due to the non-abelian Meissner effect which renders the asymptotic flatness and the locally Lorentzian spacetimes. Similarly we design for the dynamical phase, the energy momentum connection $e_{\mu}^i$ from which we have the energy-momentum form $\tau_{\mu\nu}^i$. This we interpret as the QFT curvature of the vector bundle associated with the SU(5)= SU(3)$\otimes$SU(2)$\otimes$U(1) gauge.
We then have 
\begin{equation}
 G_{\mu\nu}^i=-k\tau_{\mu\nu}^i
\end{equation}
where k is a constant that carries the appropriate dimensions in G,$\hbar$ and c. In the new framework what is the interpretation of these new field equations?  Imagine a Geometric ensemble- an ensemble of points and lines connecting those points and with each point associated a future light cone soldered to it. Then equation (1) is a wave equation wherein vibrating lattice of points and lines produce geometric ``phonons'' which are elements of SU(5)= SU(3)$\otimes$SU(2)$\otimes$U(1). These are nothing but strings. Consider a trough containing a liquid. This liquid has a latticial structure and produces acoustic phonons. The velocity limit on these phonons is that of sound which is a massless mode in that medium. similarly the velocity limit for our geometric phonons or strings is that of a photon in that medium which is spacetime wherein photon is as we all know,a massless mode. The extra dimensions can exist independent of this spacetime since the grannular structure of space is below the Planck Length and the string length exceeds this cutoff length for the ratio of the Planck length to the string length is the so called string charge which delivers a value of the fundamental electric charge which is much lesser than 1 ($=1.6\times10^-19$).

\section{Incorporating time in quantum gravity}
Consider the Hamiltonian phase space of gravity. Let us imagine our space as a fluid consisting of nodes and links. The wave function of any object can now be interpreted as the wave function of the fluid particle in the entire body of the space fluid. In the former case the wave function is represented by 
\begin{equation}
 \psi=e^{i(px-Et)}
\end{equation}
For a connection $K_{\mu}^{i}$ fix a ket $|\psi\rangle$ and an azimuthal bra $\langle\varphi|$. Then for the constraint surface representing the fluid particle, the product $\langle\varphi|\psi\rangle$ is $\psi(K,\varphi)$ Now fix an isomorphism between the constraint surface of the fluid particle and that of its bulk. This is the velocity 'v' of the fluid particle relative to the bulk of the fluid which is space. The isomorphism allows for incorporation of the Einstein velocity addition theorem. Then define a vorticity of 'v' as
\begin{equation}
 w:=vor v= curl v
\end{equation}
Now fix an automorphism in the configuration space of the bulk of space fluid. This we take as its 3-velocity 'u'.Now let
\begin{equation}
\psi(K,\varphi) = e^{i(J_z\varphi-m\Gamma)} 
\end{equation}
where $\Gamma$ is the circulation of the fluid particle. The explicit analytical continuity between the representations of eq(4) and eq(6) may then be interpreted as
\begin{equation}
 Et=m\Gamma
\end{equation}
\begin{equation}
 t=\frac{\Gamma}{c^{2}}
\end{equation}
since E=m$c^2$. Thus time arises naturally as a kind of vorticity of the fluid particle. The vorticity 'w' satisfies an equation for an inviscid liquid. For the canonical time, $\theta$ and the bulk velocity 'u' of the space, this equation reads
\begin{equation}
 \frac{\partial w}{\partial\theta}= curl u\wedge w
\end{equation}
Eq(9) tells us that the vortex lines are dragged with the fluid moving with a constant velocity 'u'.
Thus we perceive spacetime rather than space and time separately. In addition to this vorticity produces expansion and more the 4-volume more the time in that volume and hence more the expansion.
Expansion of the universe is thus directly proportional to its 4-volume.
\section{Phase Transition: Emergence of spacetime and freezing out of gravity}
Consider a gas of nodes and links henceforth referred to as grannulons for convenience. Let $E\geq E_Pl$ be the energy of the system of grannulons. The temperature of the ensemble is $T_G=\frac{\partial\sigma}{\partial E}^{-1}$ where $\sigma$ is entropy of the system. If the temperature of the gas is decreased, the gas condenses to form a liquid of grannulons. The partition function for this gas is
\begin{equation}
 z=Tr e^{-\beta[E_r-\mu N_r]}
\end{equation}
where $\mu$ is the chemical potential of the grannulons. We choose the cosmic time '$\tau$' as the order parameter. The partition function (10) changes in response to a change in energy as
\begin{equation}
 \tau= \frac{\partial}{\partial E}lnz
\end{equation}
The entropy and action are related in our context, see for example [1].
A discontinuity in the action 'S', associated with the path integral evolution of the wavefunctional on $\mathcal{C}$, given by
\begin{equation}
 S=(T\frac{\partial P}{\partial T}-P)
\end{equation}
can be expressed by the gap in the entropy $'\sigma'$ as
\begin{equation}
 disc \sigma= T_c disc\frac{\partial P}{\partial T}
\end{equation}
From (11) with $lnz = V(P-S_0)$ where $S_0$ is the ground or pre-geometric action. To this corresponds $\langle 0|\tau|0\rangle$; it follows that
\begin{equation}
 \tau=\frac{\partial S_0}{\partial E} -\frac{\partial P}{\partial E} = \langle0|\tau|0\rangle-\frac{\partial P}{\partial E} 
\end{equation}
If we expand the pressure in the vicinity of $T_c$ according to
\begin{equation}
 P=P_c+(T-T_c)\frac{\partial P}{\partial T}|T_c+...
\end{equation}
P depends on E via $P_c$ and $T_c$ as
\begin{equation}
 \frac{\partial P}{\partial E} =\frac{\partial P_c}{\partial E}-\frac{\partial T_c}{\partial E} \frac{\partial P}{\partial T}|T=T_c
\end{equation}
Inserting (16) into (14) and applying ``disc'' on both the sides, the result is
\begin{equation}
 disc \tau=\frac{\partial T_c}{\partial E} disc \frac{\partial P}{\partial T}|T=T_c
\end{equation}
While P is discontinuous at $T_c$ (as in the case of a first order phase transition),     $\frac{\partial P}{\partial T}$ may jump. From (13) and (17), we finally obtain
\begin{equation}
 \frac{\partial T_c}{\partial E}= T_c \frac{disc \tau}{disc S}
\end{equation}
Since time as an order parameter is implicitly ordering all dynamics by generating contact transformations in the phase space, eq (18) defines a chiral condensate. This provides a background spacetime for the second order phase transition in the liquid. Here we take spin as the order parameter. Now the black hole energy is written as 
\begin{equation}
 dU= \frac{\kappa c^2}{8\pi G} dA + \Omega dJ + \phi dQ
\end{equation}
We build the gravitational Gibbs Function as
\begin{equation}
 G = U-\kappa A- \Omega J+ \phi Q
\end{equation}
so that 
\begin{equation}
 dG = -A d\kappa\cdot\frac{c^2}{8 \pi G} - J d\Omega - Q d\phi 
\end{equation}
Some elementary partial differential manipulation gives 
\begin{equation}
 \frac{\partial \kappa}{\partial \Omega}|\phi_c = \frac{8 \pi G}{c^2} \frac{J_n-J_s}{A_n-A_s} 
\end{equation} 
which is identical with the Clausius-Clapeyron equation. A parallel generalisation on the basis of the work of Unruh [2] allows for $\kappa$ being replaced by the acceleration 'a' so that we have
\begin{equation}
 \frac{\partial a}{\partial \Omega}|\phi_c = \frac{8 \pi G}{c^2} \frac{J_n-J_s}{A_n-A_s} 
\end{equation}
Thus all the gravitational lines of force get pushed into the ``energy momentum bags'' and a phase analogous to the QCD confinement phase occurs at critical electric potential. The local expulsion of gravity in Lorentzian frames leads to the Strong Principle of Equivalence. Thus we have locally Lorentzian spacetime as a SU(2) superconductor and the straight line geodesics are the gravitational supercurrents in the spacetime. The heavy energy momentum bags orient the SU(2) dipoles close to them as in the case of black holes wherein the future ``dipoles'' are trapped within the event horizon. 
\section{References}
[1] S. Ohmohundro, Geometric perturbation theory in physics, (World Scientific).

[2] Unruh, W.G., "Notes on black hole evaporation", Phys. Rev. D 14, 870 (1976);Robert M. Wald (1994), Quantum Field Theory in Curved Spacetime and Black Hole Thermodynamics, (University of Chicago Press) Chapter 5;R. Mueller, Decay of accelerated particles, Phys. Rev. D 56, 953-960 (1997) ar?iv:hep-th/9706016.S.A. Fulling, Phys. Rev. D7, 2850 (1973); P.C.W. Davies, J. Phys. A8, 609 (1975).
\end{document}